 \documentclass[final,5p,times,twocolumn,authoryear]{elsarticle}

\usepackage{amssymb}
\usepackage{lipsum}
\usepackage{graphicx} 
\usepackage{grffile}
\journal{arXiv}

\begin{document}

\begin{frontmatter}



\title{Zero Trust Implementation in the Emerging
Technologies Era: Survey}

%\tnotetext[label1]{}
 
\author [first]{Abraham Itzhak Weinberg}
\affiliation[first]{organization={AI-WEINBERG},%Department and Organization
         %   addressline={}, 
            city={Tel Aviv},
          %  postcode={}, 
            country={Israel, aviw2010@gmail.com}
            %email={aviw2010@gmail.com}
            }
            \author[second]{Kelly Cohen}
\affiliation[second]{organization={Department of Aerospace Engineering, University of Cincinnati},%Department and Organization
           addressline={OH}, 
            city={Cincinnati},
            postcode={45231}, 
            country={USA}
           }

\begin{abstract}
This paper presents a comprehensive analysis of the shift from the traditional perimeter model of security to the Zero Trust (ZT) framework, emphasizing the key points in the transition and the practical application of ZT. It outlines the differences between ZT policies and legacy security policies, along with the significant events that have impacted the evolution of ZT. Additionally, the paper explores the potential impacts of emerging technologies, such as Artificial Intelligence (AI) and quantum computing, on the policy and implementation of ZT. The study thoroughly examines how AI can enhance ZT by utilizing Machine Learning (ML) algorithms to analyze patterns, detect anomalies, and predict threats, thereby improving real-time decision-making processes. Furthermore, the paper demonstrates how a chaos theory-based approach, in conjunction with other technologies like eXtended Detection and Response (XDR), can effectively mitigate cyberattacks. As quantum computing presents new challenges to ZT and cybersecurity as a whole, the paper delves into the intricacies of ZT migration, automation, and orchestration, addressing the complexities associated with these aspects. Finally, the paper provides a best practice approach for the seamless implementation of ZT in organizations, laying out the proposed guidelines to facilitate organizations in their transition towards a more secure ZT model. The study aims to support organizations in successfully implementing ZT and enhancing their cybersecurity measures.
\end{abstract}




\end{frontmatter}

%\tableofcontents



\section{Introduction}
\label{introduction}
In recent years, several phenomena have been observed in which their interactions lead to the proliferation of the ZT approach. The first is an increase in the number of devices connected to the network, such as IoTs (Internet of Things) \cite{future2021Jeffrey}. The second is poised emerging technologies, such as Artificial Intelligence (AI), Generative AI (GAI), and quantum computing. The third is the rapid growth of cyberattacks, attack surfaces, and sophistication levels of attacks \cite{business2021verizon}. In addition, the COVID-19 pandemic has accelerated the adoption of cloud-based technologies as well as technologies that enable workers to work from everywhere they would like to. In addition, the White House Executive Order (WHEO) published definitions and migration steps for ZT to agencies \cite{house2021executive}. \\
In this paper, we focus on recent technological developments in ZT. Additionally, we elaborate on their impact on ZT automation orchestration and migration in the context of new emerging technologies. To the best of our knowledge, only a few ZT survey papers relate to this aspect \cite{syed2022zero} and this is the first time that the combination between them has been surveyed.\\
This paper presents the challenges, approaches, and implementation of ZT for Detection and Response (DR) layers, such as Endpoint Detection and Response (EDR), eXtended Detection and Response (XDR), and Network Detection and Response (NDR) also known as Network Traffic Analysis (NTA). As can be learned from its name, DR has two purposes for detecting and responding to security incidents that aim to bypass the End Point Protection Platforms (EPP). ZT is a crucial framework for EDR as it is one of the most exposed points to cyberspace.\\
While dealing with defense approaches, it is important to bear in mind that while the attack can be considered a success story, even after many failures, the defense must always be successful. Hence, defense capabilities must be better than potential attacks. Cybersecurity experts claim that an attack will eventually occur, and practically, the chances of 100\% are only theoretical \cite{yampolskiy2016artificial}. Risk can be mitigated by early detection and response to an attack when it occurs. ZT is expected to solve this gap. Since ZT plays a practical role in many organizations, we find in the literature design principles such as designing architecture from inside out, determination of the access needs at a preliminary stage, focusing on the required organizational outcomes, and using the EDR as a source for inspecting the log traffic \cite{wylde2021zero}. 
The ZT can help protect against Operational Technology (OT) attacks. OT attacks are designed to exploit systems that are directly on the plant floor \cite {paes2019guide}. ZT can be applied to Industrial Control Systems (ICS) to ensure that only authorized devices and users can access the system \cite{zanasi2022zero,li2022future}. For example, in the case of the 2017 NotPetya attack \cite{greenberg2018untold}, which caused millions of dollars in damage to industrial sites, ZT principles could have prevented the spread of malware by limiting lateral movement between systems. In addition, ZT can improve Remote Access Security (RAS) by enforcing strict authentication and access control policies \cite{colombo2021access,chimakurthi2020challenge}. ZT can help protect against supply chain attacks by limiting third-party vendors’ access to critical systems and data. It can also help protect against IoT based attacks by ensuring that only authorized devices can access the network \cite{dhar2021securing,samaniego2018zero}. Finally, OT attacks using open Application Programming Interface (API) can pose a serious threat to industrial control systems, allowing attackers to gain unauthorized access to OT systems and data \cite{cheh2021analyzing}. ZT model includes strict access control and authentication policies, network segmentation, and security measures that specifically target open APIs, such as using API gateways to enforce authentication and access control policies.
The implementation of a ZT can effectively complement Digital Twins (DT) configurations.
%DT create a virtual representation of a physical system that can be used for various purposes like monitoring, optimization, planning and decision making \cite{eckhart2019digital,lv2021deep}. By replicating key aspects of physical objects or systems through sensors and real-time data, DT allows for modeling and analysis of performance, enabling testing of changes before implementation in reality. To protect and segment DT, ZT principles can be applied, necessitating strong authentication, authorization, and microsegmentation of sensitive data as well as central monitoring of logs and events is crucial \cite{sellitto2021enabling}. As DT incorporate more real-time data and assume strategic roles, securing them becomes paramount to mitigate risks of tampering or cyberattacks. ZT principles ensure that only authorized access is granted. The combination of zero trust and digital twins enhances the security, resilience, and effectiveness of IT and operational systems by integrating virtual models with granular, context-aware access controls and monitoring mechanisms \cite{jagannath2022digital}.
Digital Twins (DT) create a virtual representation of a physical system, facilitating various purposes such as monitoring, optimization, planning, and decision-making \cite{eckhart2019digital,lv2021deep}. By replicating key aspects of physical objects or systems using sensors and real-time data, DT enables modeling, analysis, and testing of changes prior to real-world implementation. To ensure the protection and segmentation of DT, the application of ZT principles is essential, encompassing strong authentication, authorization, microsegmentation of sensitive data, and centralized monitoring of logs and events \cite{sellitto2021enabling}. Using DT can increase integration of real-time data and as a result ensure robust security measures. This can lead to mitigation of the risks associated with tampering and cyberattacks. ZT principles guarantee that only authorized access is granted, enhancing the security, resilience, and effectiveness of IT and operational systems by integrating virtual models with granular, context-aware access controls and monitoring mechanisms \cite{jagannath2022digital}. The integration between two can serve as a valuable test bed for simulating cyberattacks aand evaluating the effectiveness of ZT in handling such threats.

\subsection{Zero Trust History and Evolution}
\begin{figure}[!ht]
	\begin{center}
		\includegraphics
		[scale=0.5] {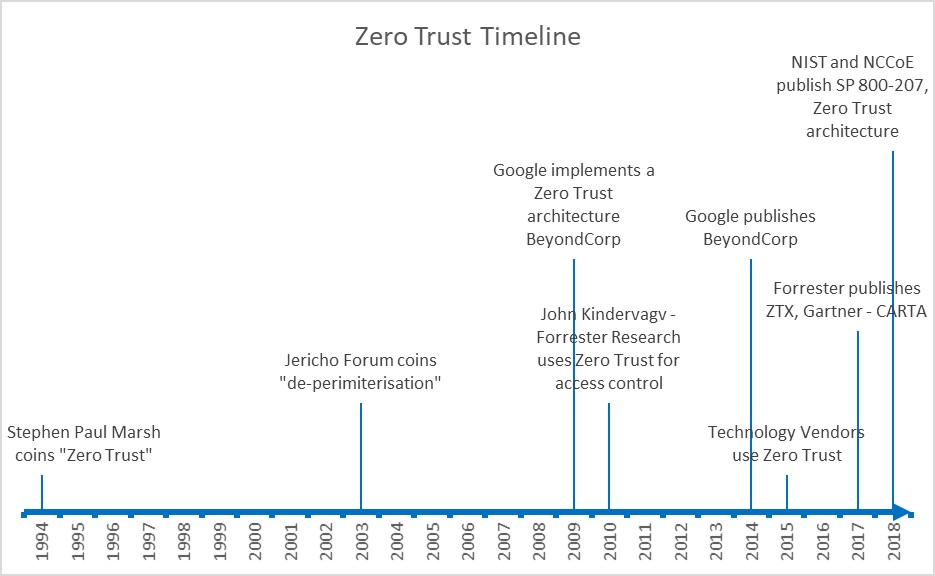}
	\end{center}
	\caption{Zero Trust Timeline}
	\label{fig:ZTTIMELINE}
\end{figure}

The phrase "Zero Trust" was coined by Stephen Paul Marsh in 1994 \cite{marsh1994formalising}. The next step occurred approximately a decade later in 2003. The Jericho Forum consortium defined de-perimeterization \cite{welborn2003jericho}. They claimed that ZT can be used as a security strategy for de-perimeterization. De-perimeterization is practically the removal of a boundary between an organization and the outside world.  
Only six years later, in 2009, Google implemented ZT architecture BeyondCorp and published it in 2014  \cite{flanigan2018zero}. BeyondCorp uses an authentication-based combination of the device and the user. Thus, it eliminates the need for privileged corporate networks.\\
In 2010, John Kindervag from Forrester used ZT for access control \cite{kindervag2010build, Forrest2010zt}. In their report, Forrester claimed that firewalls are not sufficient in the endeavor effort to cope with cyber attacks \cite{Forrest2010zt}. In addition, they coined one of the main premises principle of ZT: “never trust, always verify" \cite{Forrest2010zt}.
In 2015, ZT analysts reported augmentation in the adaptation of ZT by technology vendors \cite{alagappan2022augmenting, weforum2021ZTtimline}.\\
In 2017, Forrester and Gartner published the ZT frameworks: Zero Trust eXtended (ZTX) ecosystem \cite{cunningham2018zero} and Continuous Adaptive Risk and Trust Assessment (CARTA) \cite{campbell2020beyond} respectively. ZTX is a framework that maps ZT to organizational applications. CARTA is considered a strategic framework that enables organizations to continuously assess their risk of cyberattacks. CARTA also enables the assessment of the trust level of organizational systems. In 2018, the National Institute of Standards and Technology (NIST) and National Cybersecurity Center of Excellence (NCCoE) published cornerstone paper SP800-207. It defines cybersecurity metrics with a focus on ZT components and principles. In this way, it helps abstractly design the organizational network architecture in the light of ZT  without drilling into a specific implementation \cite{kerman2020implementing}. In Figure ~\ref{fig:ZTTIMELINE} we present the history and evolution of ZT.

Currently, ZT is widespread and has growing market potential. It is implemented in organizations worldwide that use different architectures and systems on both premise and cloud architectures. It has a prospect of 60.7 billion dollars by 2027, with an annual growth of 17.3\% \cite{ZeroTrust2027Market}. This is evident in the growth of ZT implementation and usage - horizontally all over the world and vertically - in many market segments and technologies. We can observe the invasion of ZT into 5G networks for healthcare devices \cite{chen2020security}.

\section{Zero Trust Policy and Principles}
The ZT policy is a security framework that aims to protect organizations from cyberattacks. As a result, it encompasses all the organizational systems' users, regardless of their physical location. ZT continually validates the system security configuration, as well as authentication and authorization processes \cite{stafford2020zero, rose2020zero}. \\
ZT changed the architecture of separation between different levels of security zones such as Internet, DeMilitarized Zone (DMZ), and Trusted and Privileged zones to a controlled architecture \cite{gilman2017zero, yan2020survey}.
ZT employs a Control Plane (CP) to effectively classify and distinguish between trusted and untrusted clients. When the CP acknowledges the client, it accepts the traffic. Typically, ZT uses an encrypted tunnel based on temporary one-time credentials to communicate with a trusted client. ZT transferred the traditional perimeter model and removed the borders between the different zones while maintaining secured communication only with trusted clients. \\
The Cybersecurity and Infrastructure Security Agency (CISA) proposed the ZT Maturity Model \cite{buck2021never} which divides ZT into eight pillars: users, infrastructure, devices, data, applications, networks, visibility and analytics, and orchestration and automation. The model relates to human factors, devices, and data usage. Each pillar indicates an identification process implemented before accessing the data or required services. ZT relates to both the static aspect of security that resides as data in devices and infrastructure, and to the dynamic aspect of data streams in networks and applications.\\
To achieve the ZT policy, the organization must relate to the security and authentication aspects of each pillar. Missing one of the pillars’ security aspects might influence the ZT policy and, hence, breach the organizational security level. The first and most fundamental stage of ZT is user identification. This stage is critical because most breaches use compromised identities, which are difficult to detect.\cite{kim2017certified}.\\
This stage is implemented before accessing organizational resources. ZT defines a Zero Trust Network Access (ZTNA) policy that defines which of the remote workers are eligible to connect to specific resources. In recent years, the commonly used technique for identification has been based on Multi Factor Authentication (MFA), such as Passkey. MFA is based on asking the user for two or more pieces of evidence for authentication.\\
ZT defines authorized devices and infrastructure only after going through a compliance process that stands under the Compliance Management (CM) spec. Thus, ZT ensures that the organization aligns with the required industrial cybersecurity regulations. To protect organizational data from exfiltration threats, ZT usually recommends encrypting data and using access control regulation policies, such as least privilege. ZT policy applies to applications in several ways such as legitimacy of the application, authorized user access to the application, and the environment in which the application can run. NIST defines the regulations for the ZT Application workload. Workload is a resource that supports application capability. ZT obligates organizations to define which resources are essential for each application and how they work in the most secure way.\\
To protect the network, ZT uses techniques such as explicit policies or variable trust. The variable trust uses a score to define the trust of the transferred data when an action is required. An additional ZT pillar involves visibility and analytics \cite{rose2020zero}. Visibility and analytics focuses on users and network traffic. To enable analytics and visibility, ZT guides organizations to enable network traffic inspection and store the history of logged users, asset logs, and actions. NIST ZT regulations advise the division of on-premise data centers and cloud, as well as workloads to traffic segments. Thus, ZT prevents lateral movement. In this way, the organization can investigate attacks and detect them online. \\
The eighth pillar of ZT in our list is orchestration and automation \cite{cao2022automation}. Automation and orchestration are the upper levels of ZT. They aim to deploy and apply ZT security policy. Automation focuses on making a process run without human intervention. It usually includes pipelines and scheduling parts, such as  coordinating endpoint security. Orchestration adds optimization to automation. Orchestration usually includes AI components for each pillar. It helps in detecting attacks, identifying users, and responding to cyberattacks. To the best of our knowledge, orchestration has the highest level of ZT technology.

\subsection{From legacy policies and models to Zero Trust}
ZT is considered a new approach compared to other policies and models that have been used for many years in the industry. This section discusses the relationship between them and ZT. One of the main characteristics of the ZT is its trustworthiness. To achieve this, ZT uses the Forrester's paradigm "never trust, always verify" \cite{tidjon2022never}. The journey for achieving trustworthiness from security systems started many years ago by enforcing policies and access control. Access control techniques can be divided into Attribute Based Access Control (ABAC), Role Based Access Control (RBAC), and Fine Grained Access Control (FGAC) \cite{zhou2019automatic}. The main difference between the first approaches is whether the access control depends on the access user's attributes or organizational roles. Typically, RBAC is more popular because the same user can have several organizational roles. On the other hand, FGAC uses conditions or entitlements for user access confirmation.\\
The most popular and well-known models for enforcing access control in industry are the Bell–LaPadula (BLP) and Biba. Both Biba and BLP relate to the Confidentiality, Integrity and Availability (CIA) model. CIA is designed to guide policies for information security within an organization. Biba's  main strength is to ensure the integrity of data in an organization, where BLP's ensure confidentiality. In order to achieve it Biba enforces two main rules: "no write up, no read down” \cite{xiaopeng2021zero} while
BLP enforces "write up, read down" (WURD) or "no read up, no write down". In this way, Biba's principles prevent data modification by unauthorized objects, while the BLP's rules prevent the leakage of information. To enforce the required results, both models must be implemented as Mandatory Access Control (MAC) or Mandatory Integrity Control (MIC). However, in practice, it is a challenging task to enforce the models because there are many situations that are considered grey zones. To solve these issues, Tidjon et al. \cite{tidjon2022never} proposed a ZT model that relied on and combined Biba and BLP. The authors attached trust scores to each object (such as servers, files, and data channels). The scores use weights for each object based on Biba's integrity and BLP's confidentiality levels. ZT can rely on old models and can be implemented as a top layer in organizations that have already used these approaches. 

\subsubsection{Zero Trust in comparison to legacy policies and models}
\begin{figure}[!ht]
	\begin{center}
		\includegraphics[scale=0.45]
  %[width=0.8 textwidth,natwidth=610,natheight=642]
  {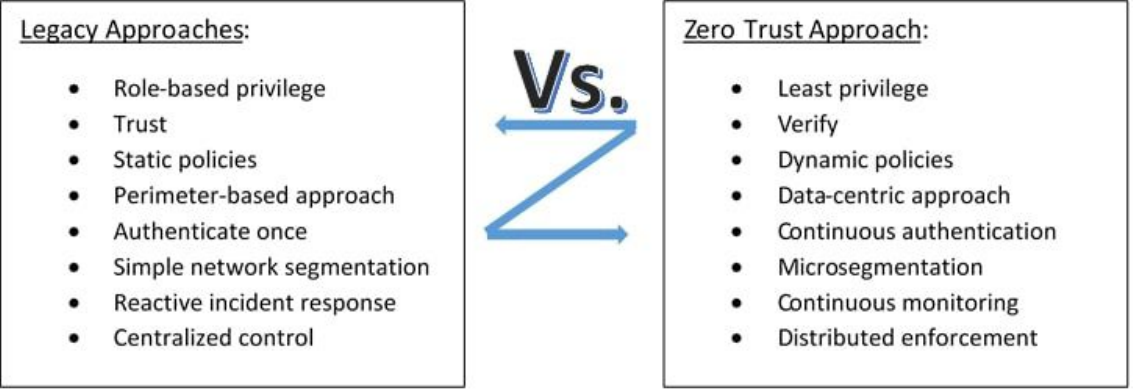}
		%[scale=0.35]
	\end{center}
	\caption{Legacy versus Zero Trust approaches}
	\label{fig:LegacyVSZT}
\end{figure}

As mentioned before, ZT is based on solid grounds of legacy policies. The adoption of ZT dictates changes in the concepts and cybersecurity worldviews. Some of the main changes are summarized in Figure~\ref{fig:LegacyVSZT}. The main differences are based on the following points: role-based versus least privilege, trust versus verification, static policies versus dynamic policies, perimeter-based versus data-centric, authenticate once versus continuous authentication, simple network segmentation versus microsegmentation, reactive incident response versus continuous monitoring, and centralized control versus distributed enforcement \cite{rousseau2021insider, jansen2022designing, syed2022zero}.\\

Legacy access control systems often rely on static roles and groups, implementing a role-based approach rather than a least-privileged one. ZT follows the principle of "least privilege”, granting only the minimum access required for a specific task or context.
Legacy models typically adopt an initial "trust but verify" approach, where devices and users are granted trust by default within the perimeter. ZT starts with "never trust, always verify" and requires continuous validation.\\
Legacy policies are typically static and undergo infrequent updates, in contrast to dynamic policies that adapt and evolve more frequently. On the other hand, ZT policies are more dynamic and situation-aware, changing based on context, such as location, task, and risk level.
Legacy models primarily prioritize securing the network perimeter, in contrast to data-centric approaches that emphasize safeguarding the data itself. Instead, ZT focuses on securing individual data assets and granting only the necessary access on a per-request basis.\\
In legacy models, users are typically authenticated once and remain trusted until they log out, as opposed to continuous authentication methods that maintain ongoing verification throughout user sessions. ZT requires the continuous authentication and re-validation of access rights.
Legacy networks approaches often exhibit coarse segmentation, with simple network segmentation practices in place, as opposed to the more granular and precise microsegmentation methods. ZT leverages fine-grained microsegmentation-based methods, for example, on data, applications, users, and tasks.\\
Legacy security practices typically rely on reactive incident response strategies that address threats after they have been detected, in contrast to the proactive and continuous monitoring approaches that provide real-time threat awareness and prevention. ZT enables proactive threat detection and containment through continuous monitoring.
Legacy security models typically depend on centralized control for policy enforcement, as opposed to distributed enforcement mechanisms that distribute policy enforcement throughout the network. ZT distributes enforcement points, such as EDR, across the full technology stack for increased resilience.

\section{Zero Trust Challenges}
ZT challenges stem from continual changes in the organizational environment. These changes can be endogenous or exogenous. Organizational management decides on internal changes and reacts to environmental changes.\\
These decisions, as well as other factors, cause organizations to constantly change workers’ positions and, as a result, their roles in the computational systems are changed. Worker status also changed on a frequent basis. New workers join organizations where others leave. In addition, each worker usually uses more than one device for his work and connects it to the organizational network. Moreover, in each device, several installed applications were used by their owners. The overall number of applications and devices is growing. An additional trend that has been evident in recent decades is data and service distributions. Organizational data and services are distributed in a wide variety of places. Some of them are located in the cloud, whereas others are on premise.\\
The implementation of ZT in organizations is a phased project. Each phase depends on the previous phase, and the final prospect results can be achieved only after years \cite{kak2022zero}. During the implementation of the ZT project, the environment constantly changes. New technologies are added, and source allocation is changed. In addition, there are many changes in budget allocations in the organization, as well as in projects and ZT levels.\\
An additional aspect of project implementation is its impact on organizational policy. As ZT is a cornerstone in organizational security, it is supposed to influence its security policy. Consequently, it can cause a domino effect on other computational and security systems, as well as attack surfaces. ZT also influences organizational standards and procedures. It is necessary to close the gap between the pre and post-implementation stages of ZT. One of the main approaches for closing this gap requires integration between departments, teams of workers, products, and legacy systems. This is usually supported by new standards and procedures. A common solution for such an integration is the creation of an organizational repository that integrates the relevant information and helps coordinate between ZT project users and leaders.\\
The ZT project requires inspecting both inside and outside the organization perimeter. Since ZT is based on the continual inspection of organizational security, continual inspection of the relevant network components is required. It requires defining tools and metrics to monitor the security status. It also has an impact on organizational risk management.
Last but not least, it requires thorough inspection of network edges such as EDR, Internet Service Provider (ISP) components, organizational IoTs, and employees’ applications and devices Bring your Own Device (BYOD). 
The challenges of ZT require not only an overall inspection of the current organizational systems but also looking outside the organization perimeter. It involves not only systems but also organizational workers and management, as well as every entity that works with the organization.\\
In the next subsections, we divide the most fundamental ZT challenges into migration, automation, and orchestration. For each challenge, we will explain how emerging technologies, such as AI, can help organizations fulfill the challenge.

\subsection{The Role of AI in Zero Trust Security}
The AI revolution is already in place. It is widely used in many fields, applications, and cybersecurity. The AI is a double-edged sword. This is evident in malware generation using GPT tools \cite{botacingpthreats}. The endless race between attackers and defenders relies on advances in AI technologies and solutions. In this section, we identify the way in which AI can integrate with ZT and elevate it to a higher level. \\
Usage in the ZT field can be classified into four main categories: CP, identity verification, attack detection, and monitoring \cite{cao2022automation}. 
The CP is the ZT brain responsible for decision making regarding whether to grant object access \cite{stafford2020zero}. \\
ZT goes hand-in-hand with the AI technology. AI can help in the implementation of ZT, as described before, and the symmetrical implementation of ZT intensifies the need for AI.
The adoption of ZT policies enhances the need to use AI based on security ecosystems. Implementing ZT in an organization creates more data that stream from more objects and sources. This also affects the microsegmentation process and intensifies the need to analyze the data at a more granular level. By implementing ZT, there is a strict need to enforce policies. AI and automation can be leveraged to dynamically define and enforce fine-grained, context-aware access policies required for a ZT model. In addition, AI can be used to continuously analyze the risk levels associated with different entities, such as users, devices, events, and processes. AI dynamically adjusts ZT policies and responses in real-time. \\ 
The continuous verification requirements of ZT make it a natural fit for AI technologies that can monitor user and device behavior in real-time and detect anomalies. This includes behavioral analytics. Rapid changes in ZT environments require fast responses to security incidents. Fast responses to security incidents that can be achieved by AI-driven automation, for instance. The implementation of ZT increases the number of system alerts. Only a small portion of alerts are relevant and must be handled. AI system reduces alert fatigue by reducing false alarms and highlighting and forwarding important issues. The nature of ZT increases the operational costs. AI enables automation of the tedious and routine operation of the ZT system, and as a result, can decrease operational costs. 

\subsection{Common AI algorithms for supporting Zero Trust cybersecuirty}
ZT classification and clustering are important techniques used in cybersecurity to identify potential security threats and protect against cyberattack. These algorithms can be divided into supervised and unsupervised algorithms, which are also known as classification and clustering. \\
Classification algorithms can be used in ZT security models for  purposes such as user and entity classification, anomaly classification, network traffic classification, file classification, and email classification. 
ML algorithms can classify users, devices, workloads, and applications into risk categories (such as high, medium, and low risk) based on their attributes and behaviors. This aids in access control and segmentation decisions. Anomaly classification can be trained based on normal behavior and activity patterns to classify new observations as normal or anomalous. This enabled the detection of threats and policy violations. Network traffic analysis can classify traffic into known categories, such as web, email, and Voice over Internet Protocol (VoIP), to enforce microsegmentation policies and detect abnormal traffic. File classification algorithms can classify files as malicious or benign to detect malware and block harmful files from entering a network. Email classifications can be used to detect phishing emails. The ML algorithms can be used to detect and filter threats before they reach the users. \\
The common classification algorithms used for ZT include Support Vector Machines (SVM) \cite{zhao2014machine}, Decision Trees (DT) \cite{el2020machine}, Linear Discriminant Analysis (LDA) \cite{alanazi2023anomaly}, Artificial Neural Networks (ANN) \cite{el2020machine}, Convolutional Neural Networks (CNN) \cite{ho2021novel}, Recurrent Neural Network (RNN) \cite{saharkhizan2020ensemble}, AutoEncoders (AE) \cite{takiddin2022deep} and Bidirectional Encoder Representations from Transformers (BERT) \cite{alaparthi2020bidirectional}.\\
Clustering algorithms can be used in ZT security models in several ways, such as device and user grouping (where the groups are not known in advance), application workload clustering, anomaly detection, threat hunting, and segmentation optimization.
Clustering algorithms can group users and devices with no prior knowledge of relevant groups. These algorithms use similar attributes, access needs, and risk profiles. This can be used to assign them to roles and enforce the least privileged access controls.
In addition, the algorithms can be used for clustering workloads based on their communication patterns, and dependencies can help optimize network microsegmentation and isolate workloads with different security requirements. Similar to classification algorithms, clustering algorithms can be used for anomaly detection where the groups are not explicit in advance. These algorithms can detect deviations in clustering patterns over time. For instance, if a node suddenly changes in a cluster, it can indicate anomalies and potentially suspicious behaviors. Threat hunting can be performed by clustering entities based on their network behaviors, and threats that exhibit different patterns can be identified and flagged for further investigation. An additional method to use clustering algorithms in the context of ZT is segmentation optimization. Network traffic patterns can be clustered to identify natural segmentation boundaries within the network, thereby aiding the design of the microsegments. \\
The commonly used algorithms for ZT clustering tasks are Affinity Propagation (AP) \cite{gao2019identification}, Density-Based Spatial Clustering of Applications with Noise (DBSCAN) \cite{wang2015adaptive}, Hierarchical clustering \cite{nishikaze2015large}, Gaussian mixture modeling (GMM) \cite{an2022ensemble,kiss2015clustering}, Self Organizing Map (SOM) \cite{kumar2021analysis} and K-Means \cite{lokhande2021trust}. \\
A combination of supervised and unsupervised learning, classification, clustering, and neural network algorithms can be leveraged to implement the core functions required for ZT security model tasks, such as anomaly detection, role-based access, and behavior analysis.

\subsection{Combating GPT and AI-Based Attacks with Zero Trust Cybersecurity Framework}
In recent years, there has been a breakthrough and profileration of AI models as well as GPT. Side by side, they enable novel attack vectors. ZT, as a cybersecurity framework, has to adapt itself to these types of attacks and provide policy and solutions for mitigating them. AI-generated attacks are a growing cybersecurity threat that can be used by malicious actors to bypass the traditional security measures. These attacks can be generated using ML models and can include a variety of tactics such as spear-phishing, malware, and social engineering. To protect against AI-generated attacks, organizations can implement a ZT model that limits access to critical systems and data, and continuously monitors potential threats. This can include strict access controls, such as MFA and role-based access controls, as well as the continuous monitoring of network traffic and user behavior.\\
Additionally, organizations can use security solutions that are specifically designed to detect and respond to AI-generated attacks, such as AI-based threat detection systems that use ML algorithms to analyze network traffic and identify suspicious activities. The chaos theory-based approach can also be used to handle AI-and GPT based attacks. 
GPT has many useful applications and can also be used maliciously by hackers to generate realistic-looking phishing emails, social engineering messages, and other types of attacks. Malware generation using GPT tools is already available in \cite{botacingpthreats}.
We dedicate the next section on how chaos theory techniques can be used to detect and respond to these attacks \cite{kumari2021performance}.
To protect against AI-generated and GPT-generated attacks, organizations can implement a ZT model that limits access to critical systems and data, and continuously monitors potential threats. Additionally, organizations can use security solutions specifically designed to detect and respond to GPT-generated attacks, such as anti-phishing solutions that use ML algorithms to analyze the content and context of emails to identify suspicious messages.

\section{Zero Trust and Chaos Theory}
Chaos theory has been used in cybersecurity applications to detect cyberattacks. Chaos theory is a branch of mathematics that studies the behavior of dynamic systems that are highly sensitive to initial conditions, meaning that small changes in the initial conditions can lead to vastly different outcomes. \\
In the context of cybersecurity, chaos theory can be used similar to AI to analyze network traffic and detect anomalous patterns that may indicate the presence of a cyberattack \cite{khan2014chaotic}. By analyzing the behavior of network traffic over time, chaos-theory-based algorithms can detect patterns that deviate from the expected behavior of the system, indicating that cyberattacks may be in progress \cite{shaukat2020chaos}.  
%One approach to using chaos theory for cyberattack detection is to model network traffic as a chaotic system and use chaos theory-based algorithms to analyze the behavior of the system. 
Additional approach is to use chaos theory-based algorithms to analyze the behavior of individual network packets and detect anomalous patterns that may indicate the presence of cyberattack \cite{alabdulkreem2023intelligent}.\\
Moreover, chaos theory can be used to compare the different states of an organization's systems and transferred data. The analysis can indicate a deviation from the normal behavior of the system, and might indicate a cyberattack \cite{okumura2021chaos}. The analyzed data can be obtained from various sources, such as logs, EDRs, networks, computers, and every device that is connected to the organization systems.\\
While chaos theory-based approaches to cyberattack detection are still in the early stages of development and are not widely used in practice, they have shown promise in detecting certain types of cyberattacks such as Distributed Denial-of-Service (DDoS) attacks. However, these approaches also have limitations such as the potential for false positives and the need for extensive training data to develop accurate models.
ZT and chaos theory can be used to enhance an organization's security posture. By combining ZT and chaos theory, organizations can implement a more effective and efficient cybersecurity strategy. As explained, ZT provides a strong security foundation by limiting access to critical systems and data and by continuously monitoring potential threats. Chaos theory analysis can be used to analyze network traffic and detect anomalous patterns that may indicate the presence of a cyberattack, providing an early warning of potential threats. 

\section{Migration Challenges and how AI can help}
ZT migration refers to the transition of an existing traditional network security model to ZT architecture. We can identify several potential challenges that organizations face when transitioning to a ZT security model, including complexity \cite{alagappan2022augmenting,lambert2022applications}, resistance to change \cite{phiayura2023comprehensive}, cost \cite{adahman2022analysis,greenwood2021applying}, visibility \cite{de2020zero}, compatibility of legacy systems \cite{teerakanok2021migrating}, integration \cite{tyler2021trust}, technology immaturity \cite{bodstrom2022strategic}, lack of skills and expertise \cite{deweaver2021exploring} and implementation time \cite{de2020zero}.
Navigating the complexities of ZT implementation requires a holistic multilayered approach that encompasses numerous changes across networks, systems, applications, processes, and policies. This level of change is inherently complex to implement \cite{alagappan2022augmenting,lambert2022applications}.
Furthermore, resistance to change can pose a challenge, as some organizations face pushback from employees who are comfortable with the status quo and are reluctant to adopt new security practices and policies. Effective communication and change management strategies are crucial \cite{phiayura2023comprehensive}. AI plays a significant role in addressing these challenges. By automatically enforcing dynamic policies, AI is well suited for implementing fine-grained, context-aware, and continually adapting access policies at the core of ZT. Additionally, AI's scalability surpasses that of humans, reducing resistance to change.\\
Another aspect to consider is the cost of implementing a comprehensive ZT model. The inclusion of various components such as MFA, microsegmentation, and continuous monitoring can be financially demanding, particularly for large organizations \cite{adahman2022analysis,greenwood2021applying}. Nevertheless, leveraging AI and ML to automate complex tasks can help lower the operational costs involved in establishing and maintaining a strict ZT architecture.
Moreover, many organizations realize a lack of visibility in their existing networks, systems, and access rights during the planning stage of ZT implementation. This lack of visibility adds further complexity to the planning and implementation processes \cite{de2020zero}. AI can contribute to improving visibility by utilizing ML techniques to analyze network traffic, endpoint data, log files, and other relevant data sources, thereby providing a clearer understanding of the existing environment. \\
In addition, the incompatibility of legacy systems with ZT features, such as continuous authentication and device posture checks, presents another challenge. Overcoming this requires the implementation of workarounds, replacements, or even the creation of air gaps to bridge the compatibility gap \cite{teerakanok2021migrating}.
Furthermore, integrating ZT solutions and technologies can be a challenging task, particularly when dealing with multiple point products from various vendors. This requires careful planning and rigorous testing to ensure successful integration. \\
In addition to the challenges of integrating ZT solutions and technologies from multiple vendors, organizations may face the hurdle of dealing with relatively immature technologies. Technologies such as identity management, microsegmentation, and other related components may still be in the early stages of development, demanding that organizations consistently update and enhance their implementation. Therefore, careful planning, rigorous testing, and a proactive approach to technology updates are essential for overcoming these obstacles and ensuring successful ZT implementation \cite{tyler2021trust}.\\
In addition, the lack of skills and expertise in areas such as identity, access management, segmentation, and cryptography can pose a hurdle for organizations adopting ZT \cite{deweaver2021exploring}. AI can alleviate this challenge by automating complex authentication, authorization, access control, and monitoring processes, thereby reducing the manual effort and training required by workers.
Implementing a full ZT transformation across an entire organization is a time-consuming endeavor, often taking months or years. The scale and complexity involved necessitate careful planning, phased implementation, and creation of roadmaps to ensure a successful and efficient transition \cite{de2020zero}.\\
In addition to the aforementioned challenges, AI can significantly contribute to ZT in various areas such as continuous verification, risk-based decisions, reducing alert fatigue, and detecting anomalies.
Continuous verification, a key tenet of ZT, can be achieved through AI technologies such as behavioral analytics and anomaly detection. These AI capabilities enable continuous monitoring of user and device activities, ensuring the ongoing verification of trustworthiness. \\
AI's role in ZT extends to risk-based decisions as it can continuously analyze risk levels associated with different users, devices, applications, network segments, and events. By dynamically adjusting ZT policies and responses in real time based on this analysis, organizations can effectively manage and mitigate risks.
AI-powered solutions have been developed to combat alert fatigue. These solutions can filter and prioritize ZT alerts, allowing human analysts to focus on critical issues. By reducing mental exhaustion and human error, organizations can maintain a high level of vigilance in their security operations.
Furthermore, AI techniques such as ML and deep learning excel to detect anomalies and outliers in large datasets. As ZT relies on continuous monitoring and verification, these AI capabilities are critical for identifying potential threats and security breaches.\\
By leveraging continuous verification, risk-based decisions, alert fatigue reduction, and anomaly detection enabled by AI, organizations can enhance the overall security of their ZT implementation, effectively mitigate risks, and ensure a robust security posture.
In summary, AI should be leveraged wherever possible during ZT migration to provide better insight, automate processes, enforce dynamic policies, continuously verify trust, make risk-based decisions, accelerate remediation, reduce costs, and detect anomalies. AI can help address many of these challenges and streamline the transition to a ZT model.

\section{Automation Challenges and how AI can help}
ZT automation uses automated technologies and processes to implement the principles of ZT architecture. Several automation challenges are associated with ZT security models. The main automation challenges revolve around integrating solutions, orchestrating policies and scaling operations, and gaining full visibility. ZT principles require the continuous verification of trust, but current technologies still rely heavily on human intervention for many tasks. Automation is an important goal; however, full autonomy remains elusive. For completeness, some of the points overlap with the migration section.\\
As mentioned above, implementing ZT architecture involves addressing various challenges in different areas. Authentication and access requests, for instance, require the automation of these processes at scale, which can be complex. This entails the integration of identity management solutions with applications, APIs, and devices to establish a comprehensive and cohesive authentication framework \cite{hatakeyama2021zero, yao2020dynamic}. \\
Similarly, the continuous monitoring of trustworthiness is crucial in ZT; however, it poses a set of challenges. The continuous monitoring of users, devices, and applications in an automated manner requires the integration of various monitoring tools to gather and analyze relevant data \cite{mehraj2020establishing}. 
On the other hand, incident response presents a challenge in fully automating the detection, containment, and remediation of security incidents, often requiring some degree of manual intervention \cite{anson2020incident, alappat2023multifactor}.\\
Device onboarding in ZT networks involves automatic provisioning of devices and applications while enforcing security policies, which can be a complex process. This requires integration of identity management, configuration management, and network access control. AI can play a crucial role in automating the provisioning of devices and applications and ensuring the enforcement of necessary security policies, configurations, patches, and updates \cite{yeoh2023zero, cheng2022universal, sanders2021integrating}.\\
Automating the issuance, renewal, and revocation of credentials such as certificates and tokens at scale is notoriously challenging in ZT. Certificate management requires careful attention to obtain this right. \cite{mehraj2020establishing}. Privileged access management is another area where automation and control of privileged access for administrative activities while maintaining operational efficiency remains an ongoing struggle \cite{devlekar2022identity}. \\
Data protection in ZT involves automatically encrypting data at rest and in transit according to security policies, which can be particularly challenging for big data and real-time systems \cite{ahmed2021improving}. In addition, automating the generation of audit reports and ensuring compliance with regulations is difficult and often relies on manual reporting, although AI can offer potential solutions in this area \cite{zakaria2019feature}.\\
Finally, API management poses a complex endeavor, especially at an enterprise scale, as it requires the automatic management of the full lifecycle of APIs, including discovery, security, monitoring, and deprecation \cite{kato2021adaptive}. These challenges emphasize the need for careful consideration, integration, and automation in multiple areas of ZT implementation to achieve a robust and effective security framework. \\
AI and ML can significantly help with automation in ZT environments. AI and ML show great potential for automating many complex components of ZT environments, such as threat detection, dynamic policy management, risk-based access control, device onboarding, accelerated remediation, and verification of trust and compliance. This can streamline the operations and reduce costs. \\
AI systems can continuously analyze the risk levels associated with different users, devices, applications, locations, and other relevant factors to make automated access control and security decisions based on risk. These capabilities are already emerging today and will likely continue to improve and expand in the coming years to further enable AI-driven automation within ZT architectures.

\section{Orchestration Challenges and how AI can help}
As mentioned above, ZT orchestration refers to the ability to automate and coordinate various systems and tools involved in organizational security. Orchestration involves integrating multiple automated systems and technologies involved in the security ecosystem, whereas automation focuses on automating individual ZT tasks and functions without direct human input. Similar to automation, the main orchestration challenges revolve around integrating point solutions, managing policies at scale, achieving visibility in operations, automating processes, and minimizing business disruptions. ZT requires coordinated actions across the full technology stack, and effective orchestration is key to success.
Some of the key orchestration challenges for ZT security models include a range of crucial aspects, such as integrating solutions \cite{bout2021machine, adamsky2018integrated}, management policies \cite{perera2022factors}, scaling operations\cite{dasawat2023cyber}, achieving visibility \cite{de2020zero}, enforcing trust \cite{qazi2022study}, automating responses \cite{eidle2017autonomic}, automated security workflows \cite{kak2022zero}, checking posture \cite{adahman2022analysis}, handling incidents \cite{sheridan2021state}, minimizing disruptions  \cite{zanasi4481853flexible}, and reducing costs \cite{adahman2022analysis}. \\
Integrating solutions from multiple vendors for tasks such as identity management, access control, device management, network segmentation, and encryption can be difficult owing to the limited interoperability among these solutions \cite{bout2021machine, adamsky2018integrated}. \\
In addition, managing various security policies across multiple systems, including access control, device configuration, encryption, and monitoring, requires complex orchestration to ensure dynamic enforcement \cite{perera2022factors}. Scaling ZT operations to support large-scale authentication, authorization, and monitoring of thousands or millions of users and devices poses further challenges that necessitate scalable solutions \cite{dasawat2023cyber}.\\
Achieving visibility by orchestrating data from various tools to obtain a unified view of users, devices, applications, networks, and threats is another challenge. Solutions often operate in silos, but AI techniques, such as ML, can analyze data from different ZT tools to provide a unified view of activities within the environment, enabling better orchestration \cite{de2020zero}. Enforcing trust and continual verification across dynamic environments, with frequent additions and removals of users, devices, applications, and networks, adds complexity to the orchestration process \cite{qazi2022study}.\\
Automating responses and coordinating solutions, such as Security Information and Event Management (SIEM), EDR, Network Access Control (NAC), and other relevant solutions in real time pose challenges, as most responses still require human input. However, AI models can be trained to autonomously integrate different ZT solutions from various vendors, minimize manual effort, and facilitate the correlation of data and alignment of policies \cite{eidle2017autonomic}.\\
Over time, AI may be able to orchestrate entire security workflows within ZT environments autonomously, from continuous monitoring to incident response, with minimal human intervention \cite{kak2022zero}. 
Another significant challenge involves effectively orchestrating security posture checks and trustworthiness assessments of devices and applications. This requires seamless coordination of operations across various solutions, including firewalls, antivirus programs, and configuration managers \cite{adahman2022analysis}.
Furthermore, orchestrating end-to-end workflows to manage security incidents from detection to resolution requires centralized visibility, control, and coordination among teams, tools, and processes \cite{sheridan2021state}. Minimizing disruptions to users and critical business processes during ZT operations requires central governance and policy enforcement for effective orchestration \cite{zanasi4481853flexible}. Finally, effectively orchestrating and automating ZT processes can optimize costs by reducing redundant tooling, streamlining purchases, and improving the overall efficiency \cite{adahman2022analysis}.\\
AI and ML can be instrumental in orchestrating ZT environments and offer significant benefits in several key areas. First, AI is well suited for dynamic policy orchestration, as it can enforce and adapt fine-grained, context-aware access policies that form the foundation of ZT. By leveraging AI, policy changes can be orchestrated in real-time across multiple systems, enabling efficient and agile policy enforcement \cite{eidle2017autonomic}.\\
AI systems can also play a pivotal role in risk-aware orchestration. Through continuous analysis of the risk levels associated with users, devices, applications, locations, and other relevant factors, AI can dynamically adjust the orchestration and prioritize them based on risk. This proactive approach allows for adaptive and intelligent decision making in managing security measures within ZT environments \cite{xiao2022sok}.\\
Furthermore, as ZT related regulations continue to evolve, AI has the potential to automate compliance orchestration. Although this remains a challenge at present, in the future, AI systems may be capable of automatically orchestrating the necessary people, processes, and technologies to ensure compliance with ZT regulations. This would streamline compliance efforts and enhance the overall effectiveness of the ZT security frameworks \cite{cao2022automation}.
The integration of AI and ML technologies into ZT environments empowers organizations with advanced capabilities to dynamically enforce policies, adapt to changing risk levels, and potentially automate compliance-related processes. The synergy between AI and ZT orchestration contributes to the overall efficacy and resilience of security measures in modern digital ecosystems.\\
AI and automation show great promise for orchestrating the complex components of ZT environments through techniques such as automated solution integration, dynamic policy management, enterprise-wide visibility, risk-aware decision-making, accelerated incident response, automated security workflows, and reduction of alert fatigue. AI can facilitate end-to-end coordination and synchronization within a ZT network.

\section{How does Quantum Computing affect Zero Trust}
Like AI, which took many years from its inception to its ubiquitous presence, quantum technology is doing its first steps in general and in the cybersecurity world and has the potential to impact ZT \cite{mogos2023quantum}. Although several quantum algorithms, such as Shor's algorithm for integer factorization \cite{de2019shor}, Grover's algorithm for database search \cite{zhang2020depth,long2001grover,lavor2003grover}, Quantum Fourier transform \cite{weinstein2001implementation,martin2020digital} and quantum counting algorithms \cite{aaronson2020quantum}, offer promising exponential or quadratic speedups in asymptotic terms, achieving practical speedup on physical quantum computers still encounters significant technological challenges. However, these algorithms do give us a glimpse of the possibilities of overcoming these challenges. As mentioned above, quantum computing remains a speculative technology. \\ 
Despite this, it has the potential to revolutionize cybersecurity in several ways. For example, it can strengthen and enable future ZT security models by providing quantum-safe cryptography, scalable policy enforcement, enhanced visibility, accelerated automation, new sensing capabilities, and more resilient network properties that are more difficult to compromise \cite{szymanski2022cyber}. However, it will likely be many years before these potentials are fully realized. The following are ways in which quantum computing could help strengthen and enable ZT security models in the future.\\
A quantum can contribute to safe cryptography. Quantum-resistant algorithms are essential for ZT solutions to withstand future attacks by quantum computers \cite{perlner2009quantum}. Technologies, such as quantum key distributions, can also enhance trust.\\
A quantum can enable the detection of anomalies. The pattern-matching and optimization capabilities of quantum computers may accelerate the detection of anomalies within ZT environments. This could help with continuous trust evaluation. Additionally, quantum computing can enhance the visibility of organizational security systems. The massive parallel processing power of quantum computers could potentially provide unprecedented levels of visibility in ZT networks, users, devices, applications, and data. Moreover, quantum technology has the potential to scale policy enforcement. In theory, quantum machines may be able to enforce fine-grained access policies required by ZT at a scale that classical computers cannot match. However, this was not possible for many years.\\
An important aspect of quantum technology is its ability to accelerate automation. The futuristic ability of quantum computers to rapidly process vast amounts of data could potentially be leveraged to automate complex ZT tasks such as continuous monitoring, threat detection, and response. When focusing on the EDR, quantum technology can be leveraged to develop sensors. New types of quantum sensors may be able to continuously monitor parameters, such as magnetic fields, vibrations, and temperatures, to help detect threats and anomalies within ZT environments. \\
Quantum has the potential to strengthen layered defenses. Quantum technologies could augment existing layers of ZT defenses, such as physical security and network segmentation, to make the overall model more resilient against threats. Finally, quantum systems are more difficult to compromise. The quantum properties of superposition and entanglement render quantum systems inherently harder to wiretap or eavesdrop on, potentially enhancing trust within ZT networks.
Quantum computing has the potential to transform cybersecurity through quantum-safe cryptography, scalable policy enforcement, enhanced visibility, accelerated automation, new sensing capabilities, and harder-to-compromise network properties. 

\section{Zero Trust Strategy and Approaches}
In this section, we describe some common ZT approaches that organizations can adopt. It discusses some of the common strategies and technologies that organizations employ as part of a comprehensive ZT approach.
The first is an identity-based access control \cite{wu2021real,wang2023secure}. This approach grants access based on verified user or device identities rather than network location. Using MFA and strong identity management systems. \\
The second method was microsegmentation \cite{sheikh2021zero,tyler2021trust}. It isolates the systems, workloads, and data within the network to minimize lateral movement if one asset is compromised. Strict control of network traffic. The least-privileged access approach ensures that users and services only have the minimal access rights necessary to perform their assigned tasks. Privileges are granted on a just-in-time and as-needed basis.
An additional relevant access approach is adaptive access \cite{teerakanok2021migrating,ahmed2020protection}. It dynamically adjusts access privileges based on real-time factors such as location, device health, and authentication strength. More stringent access controls for high-risk situations.\\
While minimizing access to entities, it is necessary to monitor organizational systems and networks on a continual basis. Continuous monitoring and analytics enables this feature \cite{stafford2020zero}. It constantly monitors network traffic, assets, and user behavior for anomalies that could indicate threats. AI and ML were used to detect deviations from normal patterns. To continually monitor organizational data traffic, gateways and service edges must be secured and monitored. Secure service edges enable the strict control of access to APIs and other external touchpoints \cite{chen2020security,stafford2020zero}. Verification of the identity, permissions, and context of each request. In addition to implementing continuous monitoring, the organization can implement User and Entity Behavior Analytics (UEBA) \cite{mehraj2020establishing,kerman2020implementing}. It plays an important role in continuous monitoring. UEBA solutions can detect anomalies and deviations in things, such as user login and access patterns, device usage and communication patterns, privileged account activity API, and application usage.
The other security aspect is related to data.\\
ZT requires data encryption \cite{syed2022zero,patil2020design}. Encrypting data at rest and in transit to prevent unauthorized access, even if the system is compromised. Therefore, key management and robust policies are critical. In recent years, there has been a transformation from an on-premise infrastructure and services to the cloud. ZT strategy involves securing a cloud workload \cite{garbis2021zero,stafford2020zero,yan2020survey}. Applying ZT principles to Infrastructure as a Service (IaaS), Platform as a Service (PaaS), and Software as a Service (SaaS) offers. Verifying managed identities, encrypting data, and enabling activity monitoring.\\
The aforementioned common approaches and technologies are part of a comprehensive, general ZT strategy. It is important to implement multiple types of controls instead of relying on a single technique in isolation. 
An additional approach to ZT involves the training and education of teams and users. Training and education are crucial for creating ZT environments \cite{vang2023factors}. Users must understand why policies have changed and how their workflows may have been impacted. This could reduce friction and encourage adoption.
From the perspective of implementation and project management, the organization must consider that ZT extends beyond technology. It also requires changes in processes, governance, and organizational culture. These include establishing a ZT Center of Excellence (ZTX CoE) and creating a ZT roadmap and milestones \cite{kerman2020implementing,stafford2020zero}.
This section described some of the common strategies and technologies that organizations employ as part of a comprehensive ZT approach. Implementing multiple types of control instead of relying on any single technique in isolation is crucial. Moreover, it is essential to remember that ZT is an ongoing journey rather than a final destination. Organizations must consistently evaluate and optimize their ZT posture in response to evolving threats and improving security maturity.

\section{Zero Trust Implementation Best Practice}
Earlier, we described the ZT migration. ZT implementation establishes a ZT posture from the start of a new environment. ZT migration transitions from an existing traditional security model to ZT architecture over time. \\
Organizations can make a smooth transition to ZT security over time by using several principles. 
To successfully adopt a ZT model, organizations should begin with small-scale pilots to gain experience, validate the value proposition, and troubleshoot any issues before gradually expanding \cite{chuan2020implementation,kerman2020implementing,campbell2020beyond,decusatis2016implementing}. The management and leading team should conduct a comprehensive inventory of all assets, including users, applications, devices, and data to establish visibility and a baseline for the ZT model. \\
To achieve the desired end state, a clear vision and roadmap should be articulated and broken down into manageable phases and priorities. It is equally important to focus on changes in security policies, procedures, and people, as it is on technologies to ensure buy-in from all stakeholders. Automation should be used wherever possible to minimize manual effort and human error, starting with easy wins to build momentum and to validate the concept \cite{buck2021never}. \\
Incorporating API-based access for applications can help minimize fragile and centralized components. A risk-based approach should be adopted, prioritizing high-risk users, assets, and access for initial ZT implementations to maximize the impact \cite{d2021building}. Careful planning and testing should be performed when integrating various zero-trust solutions to avoid these issues. A clear change management plan should be developed, communicating changes to employees, predicting impacts, and formulating a plan to minimize disruptions while providing adequate training \cite{bobbert2022zero,scott2018zero}.\\
A governance model should be created, defining roles, processes, policies, and guidelines around ZT to manage and maintain it in the long term. Regular progress measurement using metrics can help evaluate the effectiveness and identify areas for improvement, allowing for necessary adjustments. Finally, a multi-year roadmap should be developed to guide ZT implementation and achieve the desired end state over time \cite{decusatis2016implementing}.\\
In addition, it is always beneficial to follow the best practice guidelines that have already been tried in other organizations. The first principle is to start a small think-big. It is recommended that small-scale pilots start to prove value and work out issues before expanding gradually to the entire organization. Conducting a full inventory of all assets, users, devices, applications, networks, and data provides visibility and a baseline for ZT models.\\
Creating a clear vision and roadmap for the desired end state, breaking it down into actionable phases and priorities, helps organizations achieve their goals and minimizes risk and complexity.

\section {Zero Trust and Cybersecurity Technologies for Detection, Prevention, and Response}
In the previous sections, we mentioned algorithms and an AI approach for supporting the ZT response to attacks. In this section, we discuss the interaction between ZT and technologies, tools, and approaches that provide security solutions for detection, prevention, and response to cyber-attacks. ZT uses and dictates policy for tools and technologies that help organizations interact with the outside world. The definition of ZT provides a comprehensive approach to cybersecurity, helping organizations protect their critical systems and data from cyber threats. By implementing strict access controls, continuous monitoring, and risk-based authentication, organizations can reduce the attack surface and better protect themselves against potential threats. Additionally, the use of detection, prevention, and response tools can provide early warnings of potential threats and enable organizations to respond quickly and effectively to security incidents.\\
This section provides insight into the interaction between ZT and technologies and solutions such as XDR, EDR, NDR/NTA, Security Orchestration Automation and Response (SOAR) and SIEM \cite{selim2021anomaly, george2021xdr}. The section explains why the aforementioned technologies can work synergistically with ZT and enhance the organizational security level. \\
The combination of XDR can countermeasure attacks, such as Living Off The Land (LOTL) \cite{hassan2020tactical}.
XDR is a security approach that aims to integrate multiple security solutions, such as EDR, NDR/NTA, and SIEM into a unified platform \cite{george2021xdr}.\\
AI can be a meaningful tool for enhancing XDR and ZT. XDR is based on large volumes of backlogs. Even a small proportion of false alarms can cause tedious and intensive work for users. An ML-based system can decrease the human labor involved in analyzing a large amount of data. In addition, ML systems can increase the level of accuracy and, hence, decrease the number of false alarms. 
Combining the ZT policy with the XDR can practically decrease the number of logs by constantly verifying the endpoint online. However, constant verification can slow down the system response. Practically, there is a need to determine the optimal point between the response time and the required threat level. This is a practical compromise between absolute never trust, ZT policy, and organizational needs. \\
Because there is a high probability that an organization will become a target for cyberattack, one of the effective ways to handle it is by detecting it as soon as possible \cite{yampolskiy2016artificial}. Once an attack is detected, a rapid response is required. The effectiveness of an organization in handling an attack is one of the crucial parameters for its success in countermeasuring the attack. Detecting an attack at the endpoints usually indicates the early stage of the attack. Hence, implementing ZT in endpoint devices helps the organization cope with severe stages of attack. \\
XDR tools are currently attempting to bring together all the relevant security solutions. These are intended to unify multiple security capabilities into a single solution that offers automated analysis, remediation, monitoring, and detection. The organization aims to maximize the detection accuracy while increasing the security operations and remediation efficiency \cite{ali2022maturity}. 
The benefits of XDR are deemed to be so broad that Gartner called XDR the top security trend to emerge out of 2020. XDR plays a central role in advancing ZT architecture when used together with more targeted Identity and Access Management (IAM) \cite{decusatis2016implementing}.
XDR solutions offer in-depth security monitoring via flexible as-a-service delivery, which addresses identity and data monitoring \cite{ali2022maturity}. \\
As part of XDR, NDR/NTA technologies play a crucial role in monitoring and analyzing network traffic, providing insights into potential threats, and enabling rapid responses within a ZT framework.
Another important technology on which XDR relies is EDR. This refers to a category of cybersecurity solutions designed to detect, investigate, and respond to threats at the endpoint level. As part of the ZT architecture, EDR can significantly enhance the security of endpoints and strengthen the overall effectiveness of the ZT model.\\
An additional solution that enhances the ZT functionality is SOAR. SOAR is a security solution that uses automation and orchestration to streamline security operations and improve the incident response times. In the previous section, we elaborated on the relation between and ZT to automation and orchestration. 
By combining ZT and SOAR, organizations can implement a more effective and efficient cybersecurity strategy. ZT provides a strong security foundation by limiting access to critical systems and data and by continuously monitoring potential threats. SOAR then builds on this foundation by automating security operations and incident responses, enabling faster and more effective responses to security incidents. For example, a ZT model may limit access to a critical system to only authorized users and devices, and continuously monitor the system for potential threats. If a threat is detected, SOAR can automatically trigger a response, such as isolating the system or blocking access until the threat is resolved. This automated response can significantly reduce incident response times and minimize the impact of potential security breaches. \\
An additional solution that is commonly used in the organizational security suitcase is the SIEM.
The SIEM is a security solution that collects and analyzes security events from the IT environment to identify potential threats. By combining ZT and SIEM, organizations can implement a more effective and efficient cybersecurity strategy \cite{bertino2021services}. ZT provides a strong security foundation by limiting access to critical systems and data and by continuously monitoring potential threats \cite{selim2021anomaly}. SIEM then builds on this foundation by collecting and analyzing security events across the IT environment, providing early warnings of potential threats. For example, a ZT model may limit access to a critical system to only authorized users and devices, and continuously monitor that system for potential threats. If a threat is detected, the SIEM can alert the security teams and provide detailed information about the threat, enabling faster and more effective responses to security incidents.

\section{Zero Trust Evaluation}
\begin{figure}[!ht]
	\begin{center}
		\includegraphics
		[scale=0.75]{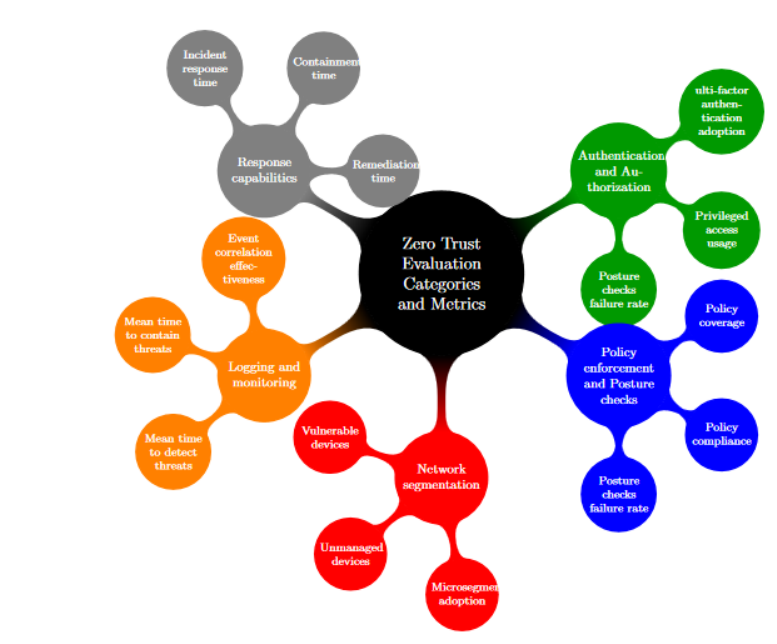}
	\end{center}
	\caption{Exampled for Zero Trust Evaluation approaches and Metrics}
	\label{fig:EvaluationMetrics}
\end{figure}

A crucial aspect of implementing ZT is to compare the organization's security state prior to and after the adoption of ZT principles. Moreover, any change in the ZT policy or architecture affects the organization. An evaluation process is required to assess the impact of these changes. The evaluation process uses metrics as a compass that enables management to understand the changes and progress while using the ZT. In this section, we present several approaches and metrics to evaluate the effectiveness of the organizational ZT security model.\\
ZT consists of several components and their integration leads to a holistic solution. As shown in  Figure ~\ref{fig:EvaluationMetrics}, the main components are policy enforcement, authentication, authorization, segmentation, monitoring, and response to threats \cite{simpson2022toward,teerakanok2021migrating}.
Policy enforcement plays a pivotal role in defining, distributing, and enforcing access policies. It is crucial to strive for consistency across identity management, authorization, and policy control points. In addition, it is important to evaluate how the ZT system enforces a least-privileged access policy for different users. Moreover, ZT uses posture checks to evaluate and  continuously verify the health of the devices accessing the network and its applications. These checks also falls into the category of XDR tools.\\
An important way for ZT to enforce its policy is to use an optimal authentication process. The evaluation of authentication must relate to all entities, such as users, devices, and applications \cite{stafford2020zero,bobbert2022zero}. Relevant metrics, such as MFA adoption, can be obtained by investigating the number of users utilizing the MFA. An additional metric relevant for authentication is privileged access usage, which can be calculated by the number of access attempts with elevated privileges. Access policy coverage can be used to identify the roles of defined access policies. In addition, access policy violations can be used to calculate violations of defined access policies. An example of a metric is policy coverage, which measures the number of assets with a defined access to configuration policies. The policy compliance metric measures assets compliant with defined access and configuration policies. An additional metric that falls in the policy category is Posture, which checks the failure rate and assesses the number of posture checks that failed to indicate non-compliance.\\
The next step in the evaluation process is authorization. One of the foundations stones is continuous authorization. The evaluation has to relate to authorization aspects such as continuity and revalidating access permissions over time. This evaluation can predict the ability of ZT to catch privilege escalations and changes. \\
An additional evaluation aspect of ZT focuses on network segmentation. This evaluation indicates how the network is segmented to restrict lateral movement \cite{basta2022towards}. Microsegmentation techniques based on application, roles, and environmental configurations can be considered. The relevant metrics for this category are microsegmentation adoption as an indication of the number of applications using microsegmentation and its complementary metric unmanaged devices, which indicates the number of devices connected without proper registration. An additional metric is a vulnerable device that assesses the number of devices with vulnerabilities identified through the posture checks mentioned above.\\
Furthermore, ZT evaluation category relates to the logging and monitoring of organizational systems. This evaluation consists of assessing organizational logs, event monitoring, and SIEM solutions. A high level of performance in this category ensures quick detection of anomalous behaviors. The metrics that can be used for evaluating the logging and monitoring categories are the mean time to detect threats and the average time to detect security incidents. The additional proposed metric is the mean time to contain threats, that is, the average time to isolate and contain security incidents. Another useful metric is event correlation effectiveness, which calculates the percentage of correlated events out of the total events.\\
The last category in our ZT evaluation process is response capabilities. It measures how quickly an organization responds to threats, revoke access, and remediate incidents. It also evaluates tabletop and red team penetration tests. Incident response metrics include the incident response time, which calculates the average time to begin the incident response after detection \cite{xiaojian2021power}. Another relevant metric is the containment time, which calculates the average time required to contain security incidents after detection. Finally, the remediation time metric calculates the average time required to fully remediate the security incidents after containment.\\
Additional qualitative metrics such as stakeholder satisfaction surveys, security assessments, and red team penetration test results can also provide valuable insights into the effectiveness of ZT initiatives.
This section provides an overview of the main ZT evaluation categories and their metrics. By evaluating these key ZT capabilities, an organization can identify gaps, weaknesses, and areas for improvement to strengthen its overall security posture. Over time, it can track progress and measure the maturity and efficacy of the ZT model.

\section{Conclusions and Future Directions}
This paper presents the main ZT issues relevant to the current era of evolving technologies. It also provides an overview of the main events that affect the ZT evolution. This study also compares the traditional ZT approaches with the current ones. In addition, it shows how organizations can implement emerging technologies and algorithms such as AI, quantum computation, and chaos theory, and how it affects the ZT strategy and implementation. Moreover, this study discusses the best practices and main issues in ZT challenges, such as migration, automation, and orchestration. Finally, this paper provides evaluation metrics and approaches for measuring the impact of ZT security on business operations, user experience, and overall security posture.
Looking ahead, we predict that the persistent and significant challenges of verification and compliance assurance will continue to evolve alongside ZT architecture. As we delve deeper into the future of ZT security, there are several critical research and development areas that warrant focused attention.
These include safeguarding and securing AI and ML workloads, establishing unified access policies across multiple domains, and implementing privacy-preserving access controls. Further, the seamless integration of ZT with DevOps processes and practices represents another promising and essential direction for future exploration and development.

\bibliographystyle{elsarticle-harv} 
%\bibliography{example}

%\bibliographystyle{IEEEtran}
%\bibliographystyle{plainnat}

\bibliography{ref.bib}
\end{document}